\documentclass[fleqn,10pt]{wlscirep}
\usepackage[utf8]{inputenc}
\usepackage[T1]{fontenc}
\usepackage{lineno}
\usepackage{multirow}
\usepackage{amsmath}
\usepackage{makecell}
\usepackage{comment}

\DeclareMathOperator*{\argmin}{argmin}

\title{C3VDv2 - Colonoscopy 3D video dataset with enhanced realism}
\author[1]{Mayank V. Golhar}
\author[1]{Lucas Sebastian Galeano Fretes}
\author[1]{Loren Ayers}
\author[2]{Venkata S. Akshintala}
\author[1]{Taylor L. Bobrow}
\author[1,*]{Nicholas J. Durr}
\affil[1]{Department of Biomedical Engineering, Johns Hopkins University, Baltimore, MD}
\affil[2]{Division of Gastroenterology and Hepatology, Johns Hopkins Medicine, Baltimore, MD}
\affil[*]{corresponding author: Nicholas J. Durr (ndurr@jhu.edu)}

\begin{abstract} 
Spatial computer vision techniques have the potential to improve the diagnostic performance of colonoscopy. However, the lack of 3D colonoscopy datasets for training and validation hinders their development. This paper introduces \textit{C3VDv2}, the second version (\textit{v2}) of the high-definition \textit{C}olonoscopy \textit{3D V}ideo \textit{D}ataset, featuring enhanced realism designed to facilitate the quantitative evaluation of 3D colon reconstruction algorithms. 192 video sequences totaling 169,371 frames were captured by imaging 60 unique, high-fidelity silicone colon phantom segments. Ground truth depth, surface normals, optical flow, occlusion, diffuse maps, six-degree-of-freedom pose, coverage map, and 3D models are provided for 169 colonoscopy videos. Eight simulated screening colonoscopy videos acquired by a gastroenterologist are provided with ground truth poses. Lastly, the dataset includes 15 videos with colon deformations for qualitative assessment. C3VDv2 emulates diverse and challenging scenarios for 3D reconstruction algorithms, including fecal debris, mucous pools, blood, debris obscuring the colonoscope lens, en-face views, and fast camera motion. The enhanced realism of C3VDv2 will allow for more robust and representative development and evaluation of 3D reconstruction algorithms.
\end{abstract}
\begin{document}

\flushbottom
\maketitle

\thispagestyle{empty}

\noindent 

\section*{Background \& Summary}

Colorectal cancer (CRC) is a significant public health concern, ranking as the second leading cause of cancer-related deaths in the United States \cite{siegel2025cancer}. Colonoscopy is the gold standard procedure for screening CRC in the gastrointestinal tract \cite{rex2015}. The timely detection and removal of precancerous lesions through routine colonoscopy screening is estimated to reduce the CRC mortality rate by 53\% \cite{zauber2012colonoscopic}. The protective value of screening colonoscopy depends on the operator’s ability to thoroughly inspect the full colon surface and identify all clinically relevant lesions. A meta-analysis of over 15,000 tandem colonoscopies revealed that approximately 26\% of adenomas go undetected \cite{zhao2019magnitude}. In a retrospective analysis, McGill et al. \cite{mcgill2021artificial} estimated that 19\% of the colon surface area typically remains unobserved during a colonoscopy, offering an explanation for a fraction of these missed lesions.

3D computer vision algorithms, including Simultaneous Localization and Mapping (SLAM), have been used to reconstruct the colon surface from colonoscopy videos \cite{ma2021rnnslam, chen2019a}. Because of the enclosed, tubular anatomy of the colon, missed regions appear as `gaps' in 3D reconstructions, and these gaps can be flagged to help clinicians provide more complete examinations \cite{ma2019}. Real-time systems employing these techniques could assist navigation during colonoscopies and increase mucosal coverage. Furthermore, advanced quality metrics derived from reconstructed topography, like fractional coverage area, can better assess screening effectiveness. 

3D reconstruction of the colon from colonoscopy video data presents several important challenges for computer vision algorithms. These include the lack of distinct features on the smooth colon surface, specular reflections, soft tissue deformation from peristalsis or distension, and occlusions caused by haustral folds \cite{ali2021deep}. Additional difficulties arise from the presence of fecal debris, rapid scope motion, limited viewpoints within the constrained tubular anatomy, a wide range of working distances present in many scenes, and non-uniform illumination from the co-located light source and camera. Despite their clinical relevance, such challenging factors are rarely captured together in existing datasets that provide quantitative 3D ground truth (GT). Thus, there is a need for datasets providing \textit{quantitative 3D ground truth for realistic and challenging colonoscopy scenarios}.

There are limited options for obtaining in~vivo colon 3D ground truth information. Current monocular colonoscopy datasets derive 3D information from Computed Tomography (CT) scans of the human colon \cite{rau2019, rau2023bimodal, rau2024simcol3d, mahmood2018}, and optical scanning of ex~vivo animal tissue \cite{ozyoruk2020}. Each approach offers a trade-off between realism and ground truth accuracy. Although datasets imaging ex~vivo \cite{ozyoruk2020} or in~vivo tissue \cite{azagra2023endomapper} provide realistic videos, they lack pixel-wise 3D ground truth. Conversely, simulated datasets generated with rendering engines offer accurate pixel-level ground truth but struggle to replicate real-world camera optics, illumination models, and light-tissue interactions \cite{mahmood2018, mahmood2018deepspie, mahmood2018deeppmb,rau2019, rau2024simcol3d, ozyoruk2020, rau2023bimodal}. Moreover, most synthetic datasets fail to incorporate realistic conditions such as fecal debris, blood, colon deformations, and the presence of endoscopy instruments. A recent synthetic dataset has made progress in this area by providing realistic vasculature, polyps, ulcers, and blood simulations, but it primarily focuses on lesion segmentation and classification, and thus lacks 3D ground truth \cite{he2025synthesized}. A summary of existing monocular colonoscopy datasets used in computer vision is presented in Table \ref{tab:comparisonMethods}.

To address these limitations, Bobrow et al. \cite{bobrow2023colonoscopy} introduced the Colonoscopy 3D Video Dataset (C3VD), which utilizes high-fidelity silicone phantoms with known 3D shapes, imaged with a high-definition clinical colonoscope. Unlike real colonoscope datasets, such as EndoSLAM \cite{ozyoruk2020} and EndoMapper \cite{azagra2023endomapper}, C3VD provides per-frame pixel-level registered ground truth data while still capturing realistic illumination, fisheye optics, and post‑processing artifacts that simulated datasets lack \cite{mahmood2018,rau2019,ozyoruk2020,rau2023bimodal}. However, the original C3VD videos only spanned idealized scenarios with clean colons and lenses, slow and smooth camera motion, a static colon, and no tools and accessories present in the field of view (FoV).

\begin{figure}[ht]
\centering
\includegraphics[width=0.8\linewidth]{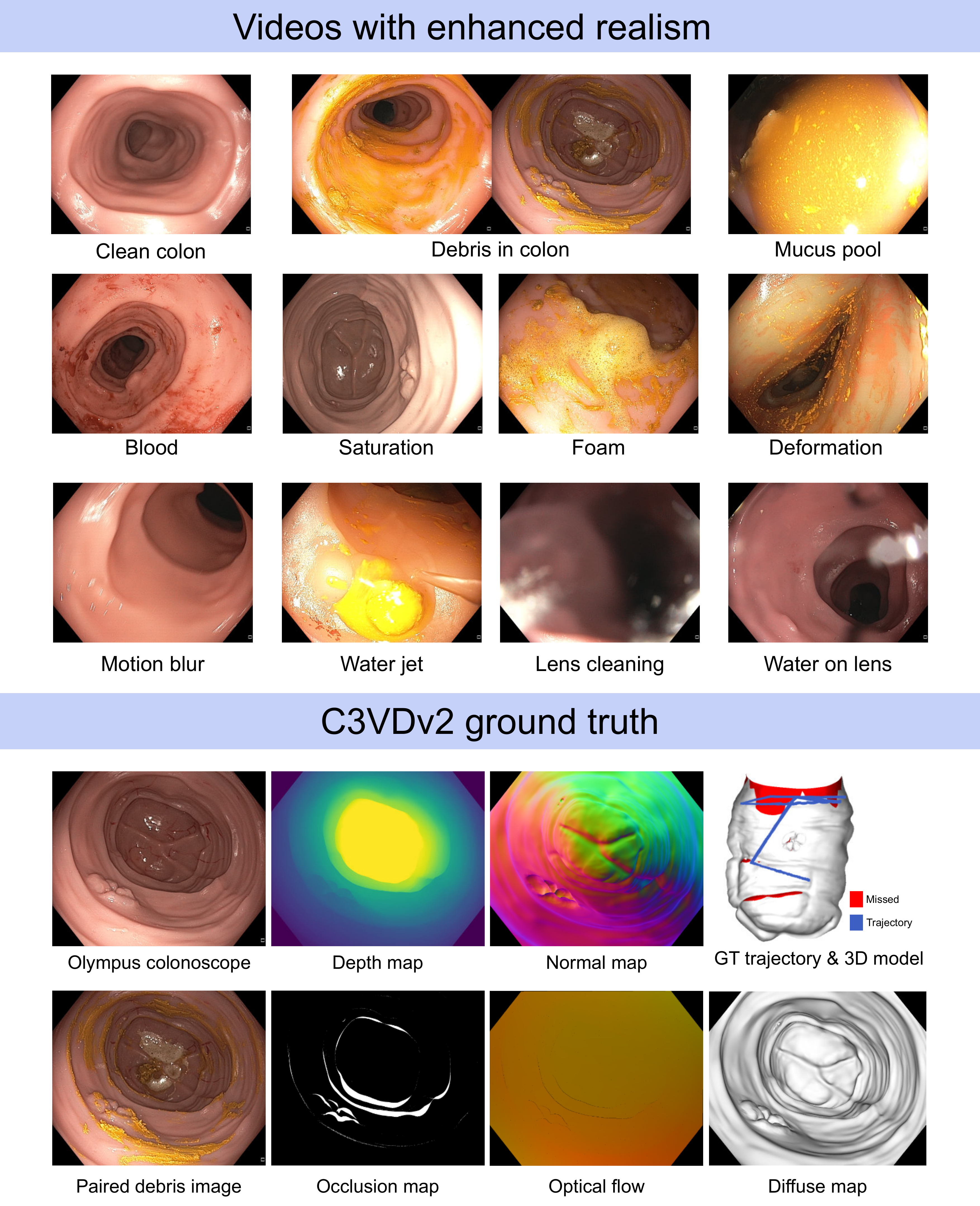}
\caption{The C3VDv2 dataset comprises 169 colonoscopy videos with paired ground truth data, including depth, normals, occlusion, optical flow, camera pose, diffuse images, and 3D model with coverage. New enhancements in the C3VDv2 dataset videos offer improved realism, adding colonic artifacts such as fecal debris, mucous pools, blood, and foam. Challenging cases like lens cleaning, water on the lens, debris on the lens, colon deformation,  and fast, non-smooth movements have been incorporated to enhance the dataset's diversity and comprehensiveness and reduce the domain gap with in vivo colonoscopy. An additional 8 videos are provided with full colon withdrawals, and 15 videos with colon deformations.}
\label{fig:overview}
\end{figure}

\begin{figure}[ht]
\centering
\includegraphics[width=0.9\linewidth]{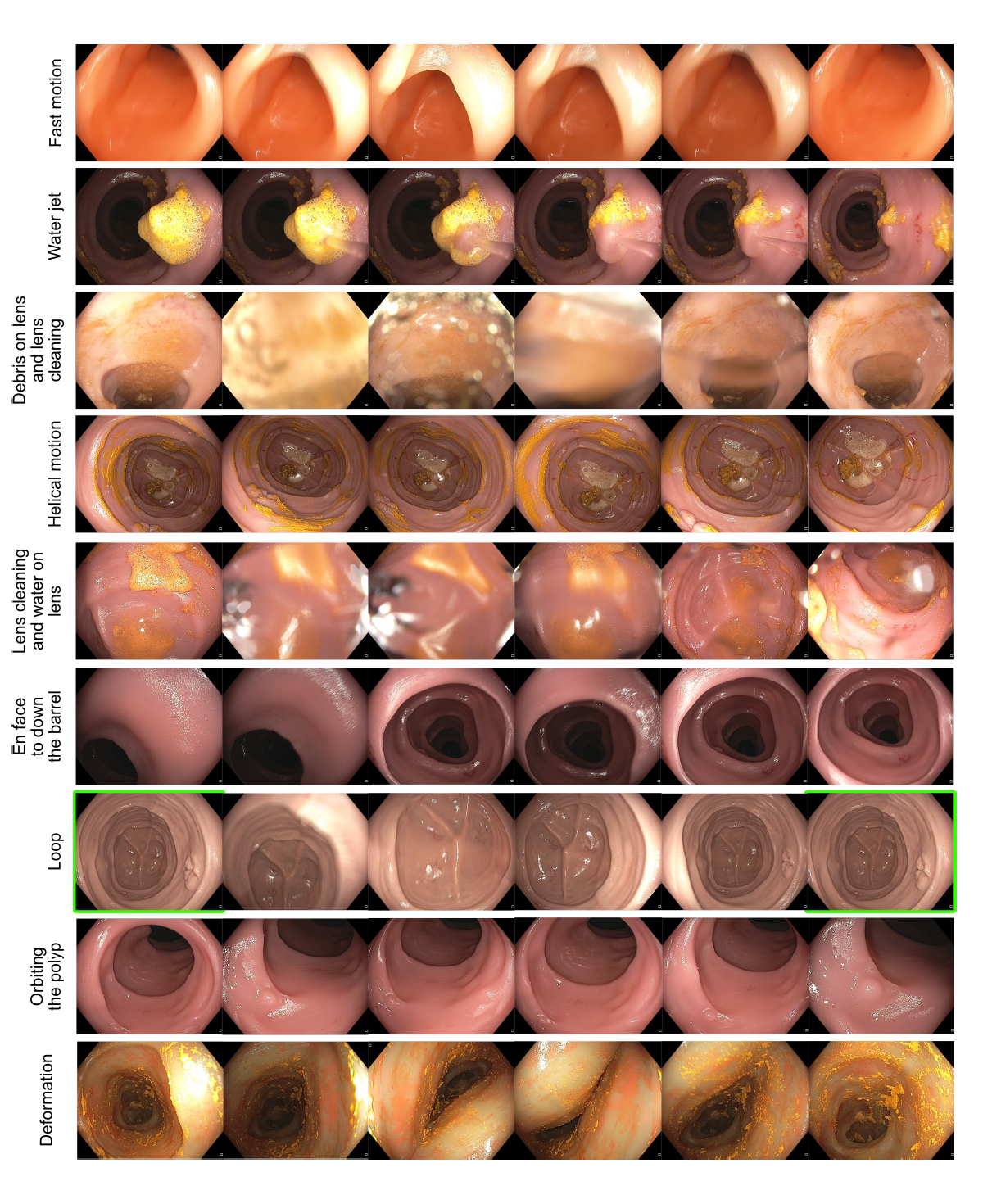}
\caption{\textbf{Video snapshots of challenging scenarios}: The C3VDv2 dataset features a wide range of challenging scenarios such as fast non-smooth motion, scope dipping in debris, lens cleaning, water on the lens, water jet, exploratory motion, loops, and en face to down the barrel transitions. Evaluating 3D reconstruction algorithms on C3VDv2 will provide a more representative assessment of their performance in real-world clinical settings. Green boxes indicate the images from the same position at the end of the loop.
}
\label{fig:Fig8_9_3_rotated}
\end{figure}

C3VDv2 retains all the advantages of the C3VD while introducing several key enhancements to improve realism, comprehensiveness, and scale:

\begin{enumerate}
	\item \textbf{Larger dataset}: The number of videos has increased sevenfold (n=192). The variety of colon geometries has doubled, with eight full colon phantoms.
	\item \textbf{Realistic artifacts}: The realism of colon appearance has been significantly improved by including artifacts such as fecal debris, mucus pools, blood, foam, and debris and water on the lens (figure \ref{fig:overview}). Furthermore, C3VDv2 incorporates instruments like water jets, lens cleaning, and suction.
	\item \textbf{Challenging scenarios}: The dataset includes scenarios that are particularly challenging for 3D reconstruction and tracking algorithms. These include fast and abrupt camera motion, en-face to down-the-barrel pose transitions, close-up en-face views of textureless surfaces, lens occlusion from debris, and lens cleaning. The dataset also includes trajectories such as straight-line in-and-out motions, loops where the first and last points are the same, and mirrored paths with lens cleaning midway (figure \ref{fig:Fig8_9_3_rotated}).
	\item \textbf{Colon deformation}: C3VDv2 contains videos of colon deformation. Due to methodological limitations, these videos lack pixel-wise ground truth maps, but they do include camera poses and the ground truth of the undeformed 3D models, making it a valuable resource for qualitative assessment or quantitative assessment after a surface transformation (“unstretching”).
	\item \textbf{Paired clean \& debris-filled colon frames}: For every debris-filled colon video, there is a corresponding clean colon video acquired by imaging a colon phantom with no debris using an identical camera trajectory as the debris-filled video. These paired videos enable the quantification of performance degradation due to debris.
\end{enumerate}

C3VDv2 is a valuable resource for the development and validation of a wide range of 3D computer vision algorithms. For example, it provides the ground truth 3D models, camera poses, and coverage meshes needed to quantitatively validate endoscopy SLAM algorithms \cite{teufel2024oneslam, elvira2024cudasift, wang2023surface}. The availability of camera poses also makes C3VDv2 suitable for volumetric rendering techniques such as Neural Radiance Fields (NeRFs) \cite{batlle2023lightneus, shan2024enerf} and 3D Gaussian Splatting \cite{bonilla2024gaussian, wang2024endogslam}. Furthermore, the pixel-level depth, surface normal, and optical flow data support the development of SLAM sub-components like depth estimation \cite{solano2025multi, paruchuri2024leveraging, recasens2021, batlle2022photometric} and camera tracking \cite{rodriguez2022tracking, rau2023bimodal}. These tasks enable downstream applications like anatomical landmark recognition \cite{taghiakbari2023automated}, improved lesion classification \cite{mahmood2018b}, polyp size estimation \cite{abdelrahim2022, iranzo2025endometric}, and augmented reality for training residents \cite{mahmud2015computer}. The dataset also includes simulated screening videos performed by an experienced gastroenterologist on full colon phantoms, providing realistic camera trajectories for validation of camera pose estimation methods. The diversity in tissue color, texture, and camera trajectories enables testing for domain generalization and robustness of these algorithms.

C3VD’s high-resolution 3D models and phantom-making protocols are open-sourced, enabling researchers to modify, 3D print colon molds, and cast their own phantoms for imaging. These protocols can be expanded to fabricate other tissue phantoms, such as other regions of the gastrointestinal tract, like the esophagus and stomach, or to create organ structures
for surgical datasets. Beyond 3D reconstruction, C3VDv2 can also be used for auxiliary colonoscopy tasks, such as artifact detection \cite{ali2020objective} and haustral fold detection \cite{golhar2025HalFSAM ,jin2023self} in the presence of debris. The dataset could also be used to generate synthetic data \cite{golhar2024gan, psychogyios2023realistic} for semi- or self-supervised learning \cite{golhar2020improving}. Researchers may further annotate bowel‑preparation quality of videos using the Boston scale \cite{cold2024development} to gauge performance across varying levels of debris. Since text labels describing artifacts and events in the videos are also provided, the video-text data could aid training medical vision-language models \cite{sharma2025diverse}. C3VDv2 also holds potential for polyp detection and tracking. Some polyps have been strategically positioned to be occluded by haustral folds. Lastly, fecal debris could generate false-positive detections, presenting a challenging validation set for lesion detection algorithms. 

In summary, C3VDv2 substantially extends the scale and realism of the original dataset, enabling comprehensive evaluation of computer vision methods under diverse and clinically relevant colonoscopy conditions.

\section*{Methods}
The C3VDv2 dataset includes colonoscopy videos with pixel-level ground truth maps. To achieve this, we first established a protocol for fabricating high-fidelity phantoms from a 3D surface model of a human colon. Then, we record video sequences of a colonoscope advanced through colon phantoms with a robotic arm while recording the pose log. Finally, we used a multimodal 2D/3D video registration technique to align the acquired video and trajectory data with a ground truth 3D surface model of the colon. A more detailed description of the registration method is described in C3VD \cite{bobrow2023colonoscopy}.

\subsection*{Silicone colon phantom fabrication}
To develop realistic colon phantoms for colonoscopy research, two complete 3D colon models were sculpted using MeshMixer software (AutoDesk, Inc.) based on colonoscopy videos and the C3VD model. Each model was reviewed by an experienced gastroenterologist (V.S.A.) for anatomical accuracy. Colon 1 (c1) was divided into eight distinct regions, while Colon 2 (c2) comprised seven regions, aligning with anatomical sections - cecum, ascending colon, transverse colon, descending colon, sigmoid colon, and rectum; further subdivisions such as transverse 1 and transverse 2 were created for longer segments to fit the 3D printer’s build volume (14.5 × 14.5 × 18.5 cm). Lesions representing various categories of the Paris Classification \cite{lambert2003paris} were sculpted onto the models and strategically placed such that the haustral folds would occlude some polyps. Comprehensive details regarding the lengths of model segments, types of lesions, and their major axis diameters are presented in Tables \ref{tab:colon1-stats} and \ref{tab:colon2-stats}, and an illustration is presented in figure \ref{fig:colon_crossection}.

Three-part molds were created for each colon segment. The molds comprised two outer parts forming the phantom's external shell and one insert part forming the colon lumen. All insert parts were 3D printed on a high-resolution Formlabs Form 3 printer using the adaptive layer thickness setting, which used a layer height between 25 and 200 micrometers to depict the colon lumen details accurately. The outer shells, which form the external surface not seen by the colonoscope, were printed using an Ultimaker S3 printer with a 0.2-millimeter resolution.

\begin{figure}[ht]
\centering
\includegraphics[width=\linewidth]{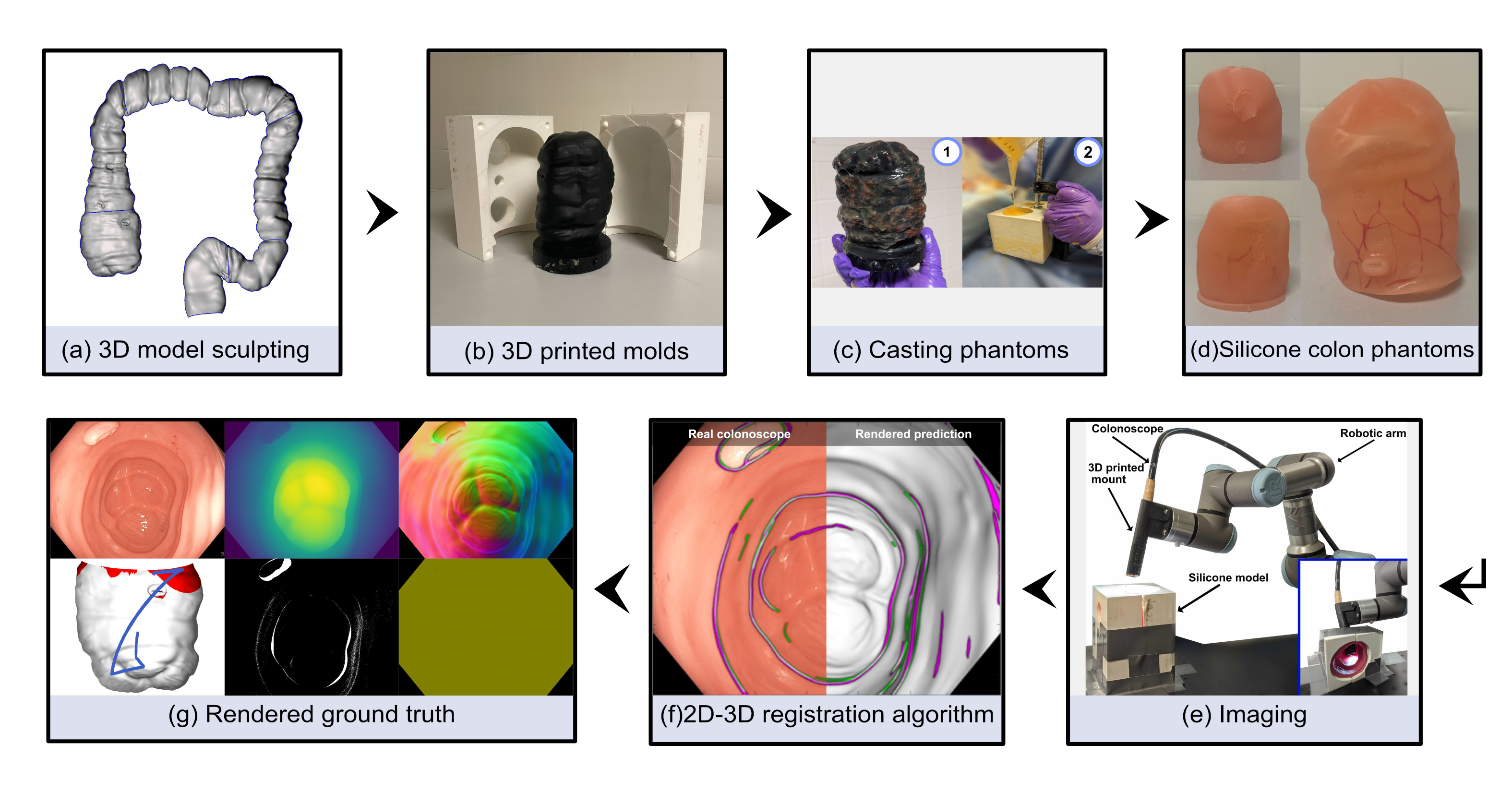}
\caption{Overview of the C3VDv2 data collection process. \textbf{(a)} High-resolution 3D models of colons are digitally sculpted. \textbf{(b)} Three-part molds are 3D printed. The high-resolution core insert (black) represents the colon lumen, while the outer shells (white) define the external contours of the phantom. \textbf{(c)} The phantom casting process involves (1) painting silicone layer-by-layer onto the insert mold with the vasculature hand-painted on intermediate layers. The painted insert is then placed in the outer shells, and (2) silicone is poured through mold inlets to fill the space between the insert and outer shell, \textbf{(d)} resulting in the final colon phantom. \textbf{(e)} The imaging protocol involves rigidly attaching the colonoscope tip to a robotic arm for precise, repeatable, and pose-tracked maneuvers. Phantoms are secured inside their outer shells to maintain shape. \textbf{(f)} 2D/3D registration is performed to align the 3D model with the 2D RGB video frames. The loss function is optimized based on the overlap of the corresponding edge maps. \textbf{(g)} After alignment, a virtual camera is navigated along the camera trajectory to render pixel-wise ground truth frames.}
\label{fig:flowchart}
\end{figure}

\begin{figure}[ht]
\centering
\includegraphics[width=\linewidth]{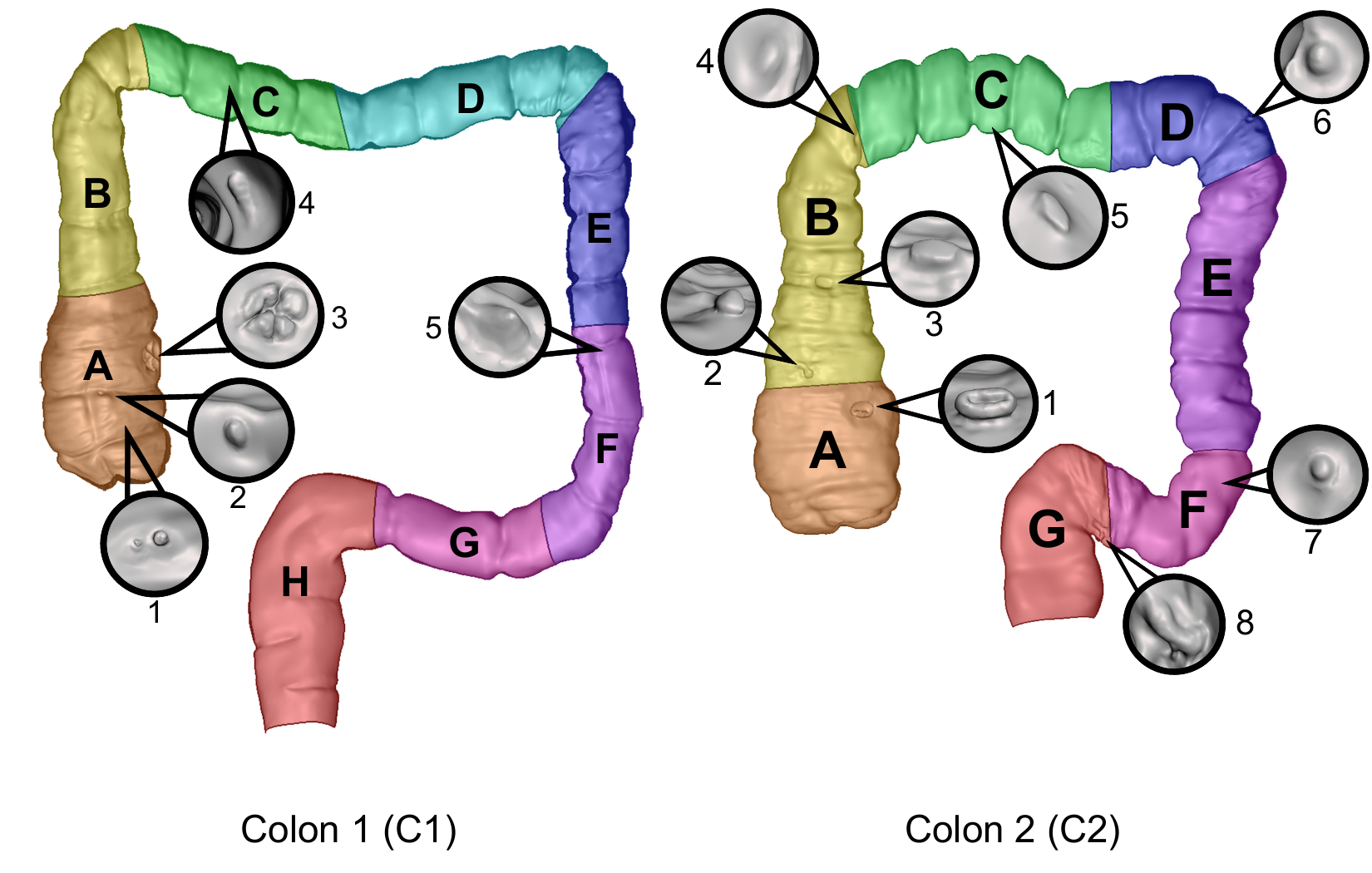}
\caption{\textbf{Cross-section of Colon models}: High-resolution 3D models were digitally sculpted referencing anatomical images. For positional reference, magnified views of lesions and Ileocecal valves are shown. The colors represent the distinct colon segments used for making phantoms. Quantitative details such as segment lengths and polyp sizes are provided in Table \ref{tab:colon1-stats} and \ref{tab:colon2-stats}.
}
\label{fig:colon_crossection}
\end{figure}

Silicone (Dragon Skin, Smooth-On, Inc.) was used to create casts from each mold, with color and texture variations introduced using silicone pigments (Silc Pig, Smooth-On, Inc.). For each phantom, the initial layers of silicone were manually painted onto the insert. 5 to 12 layers were applied to allow for detailed hand-painting of features such as vasculature patterns on intermediate layers. Silicone was later poured into the bulk of the mold through inlets in the outer molds, filling any gaps between the insert and outer shells. This entire process is illustrated in figure \ref{fig:flowchart}. 

\begin{figure}[ht]
\centering
\includegraphics[width=\linewidth]{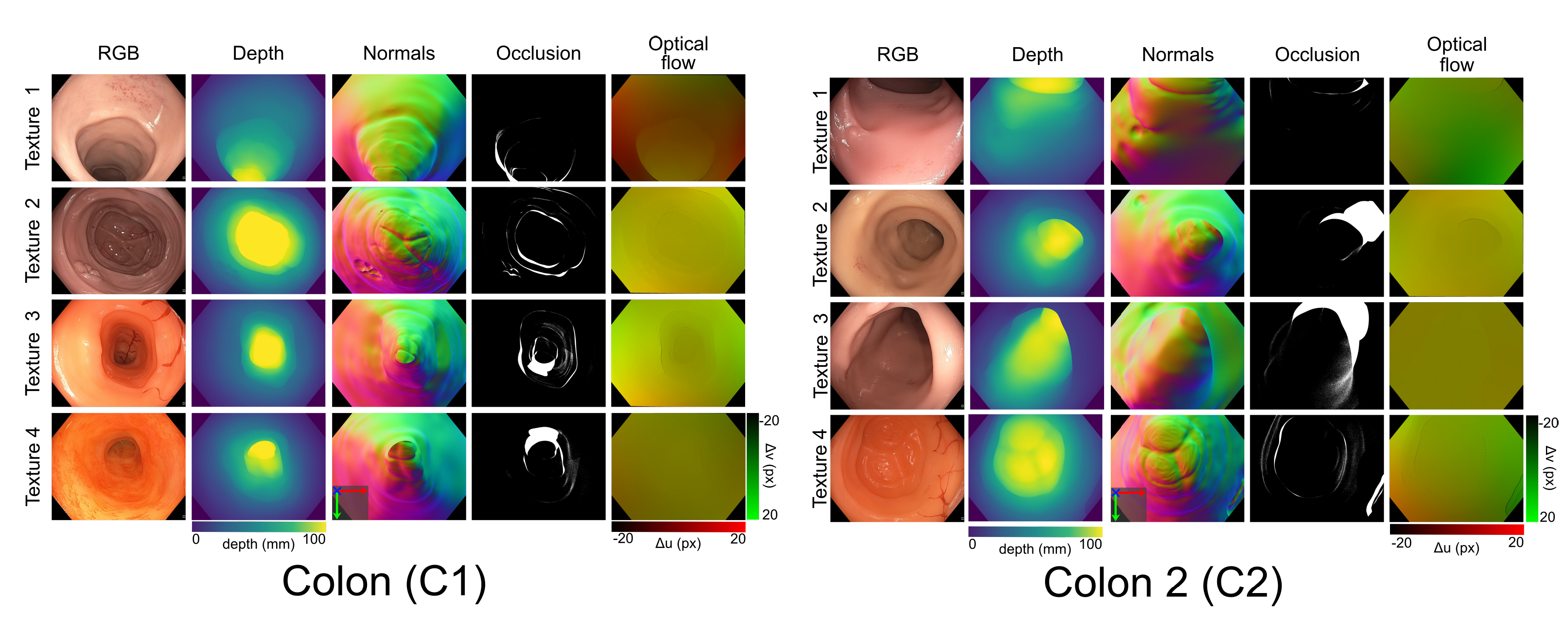}
\caption{\textbf{Sample frames from `clean’ colon sequences (Video v1 \& Video v2):} Four phantoms with varied tissue colors and textures were cast for each colon model. Paired ground truth maps are available for all RGB frames. 
}
\label{fig:clean_pairs}
\end{figure}

Using the molds for each model segment, four complete phantoms were cast, each varying in texture and color. This effort resulted in a total of eight full-length colon phantoms as shown in figure \ref{fig:clean_pairs}. The diversity in tissue colors and textures among these phantoms was explicitly designed to enable domain randomization for robust training and evaluation of computer vision algorithms. 

Future versions of these phantoms could incorporate tuned optical scattering and absorption properties to simulate more realistic light–tissue interactions \cite{ayers2008, sweer2019, chen2019b}. Additionally, phantoms embedded with polyps suitable for simulated polypectomy procedures \cite{zhao2023make} would support the integration of therapeutic endoscopy tools in the dataset.

\subsection*{Data Acquisition}

\subsubsection*{Imaging Setup}
 
The experimental setup employed for data acquisition is illustrated in figure \ref{fig:flowchart}. An Olympus CF-HQ190L video colonoscope, CV-190 video processor, and CLV-190 light source were used to image the phantom models. Uncompressed video frames were captured from the clinical video processor using an Orion HD (Matrox Imaging) graphics adapter.  Video frames were parsed and stored in raw uncompressed RGB format. Silicone models were placed inside the outer shell molds to prevent phantom motion and deformation during video recording. All videos were recorded using white light illumination. The colonoscope's distal tip was rigidly and securely mounted to the end effector of a Universal Robotics UR-3e robotic arm. 6-12 waypoints were manually programmed for each video segment to simulate colonoscopy trajectories. The robotic arm traversed the colon by interpolating between these poses, with a repeatability of $\pm0.03$ mm. A pose log was recorded from the arm at a sampling rate of 500 Hz.

\subsubsection*{Enhanced Realism}
\label{sssec-enhanced realism}
\begin{figure}[ht]
\centering
\includegraphics[width=\linewidth]{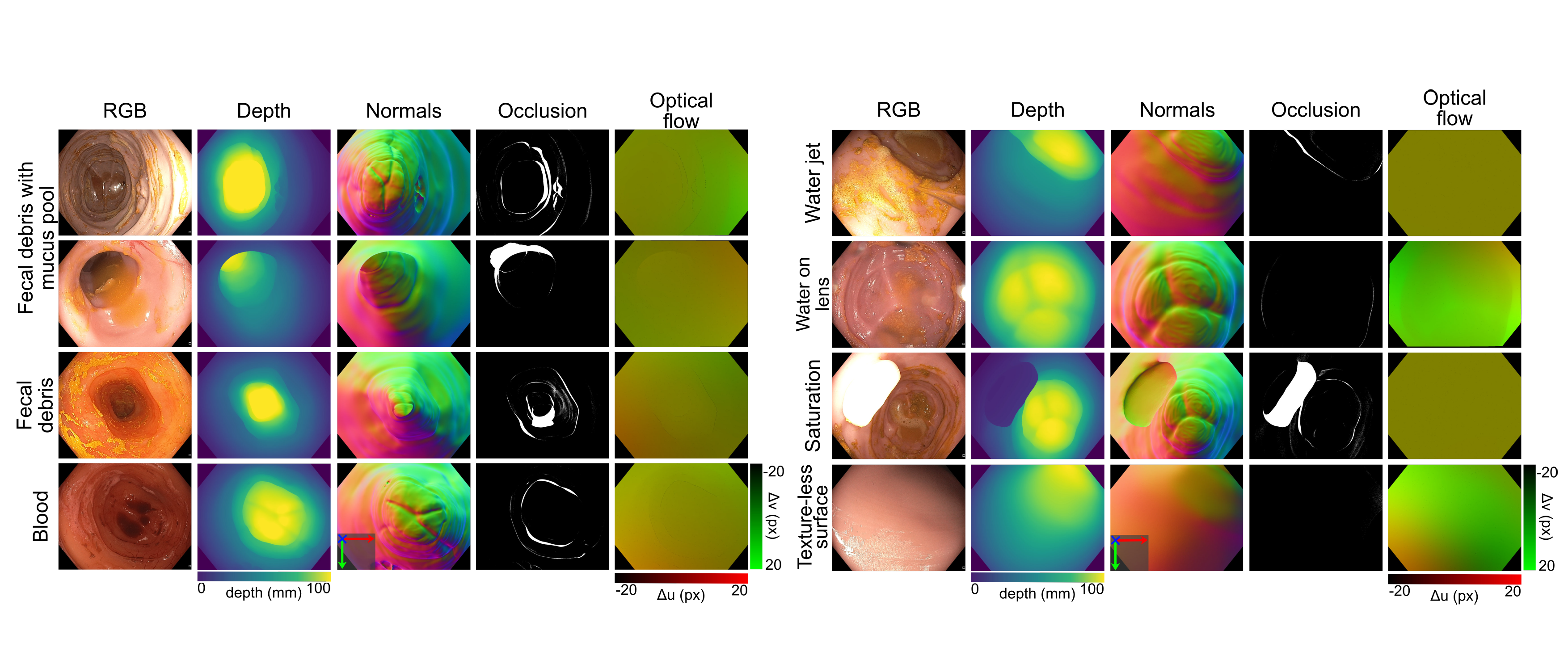}
\caption{\textbf{Sample frames from debris-filled colon sequences (Video v3):} C3VDv2 bridges the gap between phantom datasets and real-world colonoscopy by including colonic artifacts like fecal debris, mucous pools, blood, foam, en-face textureless wall views, and water on lens, which often pose difficulties for 3D reconstruction algorithms. The frame-wise ground truth depth, normal, occlusion maps, and 3D model align with the `clean' colon ground truth and do not take into account the additional debris. 
}
\label{fig:debris_pairs}
\end{figure}

Colonoscopy artifacts, such as specular reflections, mucous pools, fecal debris, blood, and foam, pose significant challenges for 3D reconstruction algorithms. These artifacts were introduced to increase the realism of the colonoscopy videos. Real colonoscopy videos served as a reference for creating the debris-filled colon environment. To emulate the specular reflections of the mucosa, the phantom was coated with a silicone lubricant (015594011516, BioFilm, Inc.). A mixture of peanut butter (051500255377, Jif), applesauce (085239045145, Good \& Gather), and yellow pigment powder (B0B82NLDDR, Rock N Soil) mimicked fecal debris. Different grades of bowel preparation were created by varying the quantities of debris. Simulated blood was created by combining glycerine (754207111907, Raw Plus Rare), red dye (B089HYVC3R, Craft County), and applesauce. Mucous pools were simulated using a mixture of glycerine, yellow dye, and chicken soup (051000134592, Campbell). Soap foam, colored with yellow pigment powder or red dye, was also added to the debris. The colonoscope's auxiliary water jet system was used to flush debris from the phantom's walls. Water-on-lens and lens cleaning scenarios were captured using the colonoscope's lens cleaning system. In simulated screening videos, the system's suction feature was used to drain mucous pools and debris. Representative examples are shown in figure \ref{fig:overview} and \ref{fig:debris_pairs}. These debris formulations were reviewed by an experienced gastroenterologist for qualitative agreement with clinical scenes.

\subsubsection*{Pixel-level GT \& Deformation Video Collection Protocol}
For each phantom, a maximum of four videos were recorded. The first and second videos (v1 \& v2) captured the clean colon, exhibiting only specular reflection and occasionally lens cleaning. These videos had different imaging settings, such as illumination intensity, and trajectories. Debris was then added to the phantom for the third video (v3), with the camera trajectory from v2 being replicated. In a few cases, a colon deformation video was also captured (v4). The phantom was removed from the molds and deformed using a blood pressure cuff (31191721394, Walgreens) inflated using an air pump (Pacum, Masterspace). The deformation video (v4) features either a stationary camera or linear camera motion.

\subsubsection*{Simulated Screening Colonoscopy}
In addition to the short videos featuring phantom segments, simulated screening colonoscopy videos of complete colon phantoms were recorded. Individual phantom segments were bonded together using a silicone adhesive (Sil-Poxy™, Smooth-On, Inc.) and selectively sutured for structural reinforcement. As discussed previously, materials such as glycerin and yellow dye were incorporated to simulate debris. The full colon phantom model was mounted on a cutout foam scaffold to reduce global movement and colon deformation during imaging.

For camera tracking in this experiment, a six-degree-of-freedom electromagnetic (EM) sensor (Aurora, Northern Digital Inc.) was affixed to the distal end of the colonoscope, and poses were sampled at 40 Hz. An experienced gastroenterologist (V.S.A.) performed the simulated screening colonoscopy to capture realistic camera trajectories. In total, eight full-colon simulated screening colonoscopy videos (withdrawal phase) were recorded, each corresponding to different texture phantoms and debris. The average duration for these videos was approximately 6.5 minutes, consistent with the >6-minute withdrawal phase duration guideline issued by the American College of Gastroenterology taskforce \cite{rex2009american}.

\subsection*{2D/3D Registration Pipeline}
The 2D/3D registration process proposed in C3VD \cite{bobrow2023colonoscopy} was followed. A virtual camera was traversed along the recorded real colonoscope trajectory to generate pixel-level ground truth for each video frame. This virtual camera rendered corresponding ground truth frames from the 3D model of the colon. A key challenge is determining the relative location of the phantom model in relation to the virtual camera's trajectory. Assuming the phantom model remained stationary during the video capture, its pose can be represented by a single, rigid body transformation. To estimate this unknown transformation, a 2D/3D registration approach was utilized. This approach aligns the virtual 3D model with the 2D video frames using extracted edge maps. The registration process iteratively samples the model transform parameter space and measures alignment between edge maps from the video frames and depth maps rendered by the virtual camera. 

\subsubsection*{Data Preprocessing}
The recorded video frames and pose log are preprocessed before the registration step. Five keyframes were uniformly sampled from each video sequence. This number was empirically chosen to balance registration accuracy and computational efficiency in C3VD \cite{bobrow2023colonoscopy}. If a sampled frame lacked informative edges (e.g., textureless regions) or exhibited artifacts (e.g., lens cleaning), a nearby frame with clearer edges was manually selected as a substitute. After selecting keyframes, preprocessing was applied to facilitate the generation of high-quality edge maps. Specular reflections were first removed using Telea inpainting \cite{telea2004image} to suppress spurious edges. Due to the diverse range of tissue colors, textures and lighting conditions present in the phantom videos, contrast enhancement was necessary. Keyframes were denoised using a Gaussian filter and converted to grayscale. Subsequently, Contrast Limited Adaptive Histogram Equalization (CLAHE) \cite{zuiderveld1994contrast} was applied to enhance local contrast. Then, edge extraction was done using the deep learning-based method DexiNed\cite{xsoria2020dexined}. The pre-trained model weights (BIPED\_10.pt) available from the DexiNed repository were used for this purpose. DexiNed edge maps were inverted (black background, white edges) to match conventional edge formats. Non-maximum suppression and hysteresis thresholding from the Canny edge detector \cite{canny1986computational} were applied to produce clean binary edge maps. These preprocessing steps generated binary edge maps with thin edges from the input RGB key frames.

The pose log was preprocessed to reduce noise. A box filter with a window size of 54 ms was applied to the data. The pose log was then downsampled to 60 Hz. It is important to note that the pose log records the position of the robotic arm's end-effector relative to its base, not the camera itself. Hence, a hand-eye calibration was performed to estimate the camera poses. The calibration computed a transformation matrix $\mathbf{X}$ relating the robotic arm motion $\mathbf{A}$ to the colonoscope camera motion $\mathbf{B}$. This relationship was expressed by the equation $\mathbf{A}\mathbf{X}=\mathbf{X}\mathbf{B}$. The Park \& Martin \cite{park1994} optimization method was used to solve for the unknown transformation $\mathbf{X}$. Synchronization between video frames and pose logs was achieved by computing the temporal offset that maximized the correlation between optical flow magnitudes (representing camera motion) and the translation component of the pose log. Finally, after transforming robotic arm poses into camera poses and synchronizing them with the video frames, the camera pose for each keyframe was computed by interpolation between the temporally closest camera poses.  

For simulated screening videos, synchronization between video frames and EM tracker poses was achieved by manually identifying temporal correspondence points, fitting a linear regression model (time\_sensor = drift × time\_video + offset) to correct for clock drift, and interpolating poses to match adjusted frame timestamps. Clock drift arises due to the lack of hardware synchronization between the colonoscope's video acquisition and the EM tracker, which operate on separate internal clocks. Over time, small differences in their clock frequencies accumulate, leading to misalignment between recorded timestamps. The linear model compensates for this temporal drift, enabling accurate alignment of pose data with corresponding video frames. After correcting timestamps, the camera poses were obtained from the EM tracker poses via hand-eye calibration as described earlier.

\subsubsection*{Fisheye Camera Model}
An essential feature of clinical colonoscopes is their wide FoV, which is typically around $170^{\circ}$. This feature maximizes the colon surface area visible during a colonoscopy procedure. C3VD incorporated this fisheye effect in its virtual camera model to render the pixel-wise ground truth. As in C3VD, the spherical omnidirectional camera model proposed in Scaramuzza et al. \cite{scaramuzza2006} was used in this dataset.

The spherical camera model establishes a one-to-one correspondence between a pixel $(u", v")$ on the image plane and a ray direction in 3D world space $(X, Y, Z)$. The Scaramuzza model employs a polynomial projection function to map the ideal undistorted image coordinates $(u, v)$ to the ray direction, by:

\begin{equation}
	\begin{bmatrix}X \\ Y \\ Z \end{bmatrix} =
	\lambda \begin{bmatrix} u \\ v \\ \alpha_0 + \alpha_2\rho^2 + \alpha_3\rho^3 + \alpha_4\rho^4 \end{bmatrix},
\end{equation}

\noindent where $\lambda$ is a scalar factor, $\alpha_i$ are the polynomial coefficients ($\alpha_1 = 0$), and $\rho= \sqrt{u^2 + v^2}$ is the radial distance of the pixel from the image center. To account for lens-sensor misalignment and distortion, the ideal image coordinates are transformed into real distorted image pixels $(u", v")$ using the following equation:

\begin{equation}
	\begin{bmatrix} u" \\ v" \end{bmatrix} =
	\begin{bmatrix} c & d \\ e & 1 \end{bmatrix}
	\begin{bmatrix} u \\ v \end{bmatrix} + 
    \begin{bmatrix} c_x \\ c_y \end{bmatrix}
\end{equation}

\noindent where $c,d,$ and $e$ are elements of the stretch matrix and $(c_x, c_y)$ correspond to center shift. These intrinsic camera parameters were computed during a camera calibration measurement. An $8 \times 11$ checkerboard with a square size of 10 mm was utilized as a calibration target.  The MATLAB 2023b Camera Calibration toolbox (Mathworks) was used to compute the fisheye camera intrinsics, which resulted in a mean reprojection error of 0.43 pixels. The camera intrinsics are provided in \textit{camera\_intrinsics.txt} file.

\subsubsection*{CMAES Optimization and Edge Loss}
The 2D/3D registration aligns geometric features between the video frames and the 3D model. The optimization and loss function proposed in the original C3VD \cite{bobrow2023colonoscopy} study were used. During this process, the virtual camera traversed the pre-defined camera trajectory, rendering depth maps from the 3D colon model at each keyframe pose. At each registration iteration, the current transformation estimate ($\mathbf{T}$) was applied to the 3D model. Depth maps were then rendered at camera poses corresponding to the selected keyframes. Canny edge detection was used to extract edge maps from the rendered depth images. An edge similarity score $\mathcal{S}_e$ was used to quantify the similarity between the edge maps obtained from the video keyframes and the rendered depth maps. After blurring the edge maps with a Gaussian kernel, the edge similarity score between the rendered depth's edge maps $E_{\text{Depth}}$ and the target edge maps extracted from video keyframes $E_{\text{RGB}}$ was computed as:

\begin{equation} \label{eqn:objective}
\mathcal{S}_e = \frac{1}{\textrm{KHW}}\sum_{k=1}^{\textrm{K}} \sum_{v=1}^{\textrm{H}} \sum_{u=1}^{\textrm{W}} E_{\text{RGB}}^k(u,v) \cdot E_{\text{Depth}}^k(u,v),
\end{equation}

\noindent where K is the number of keyframe pairs, H and W represent the image height and width, respectively, and the dot product is applied between corresponding pixels in the edge maps. The similarity score was incorporated into the optimization objective for estimating $\mathbf{T}$:

\begin{equation} 
\mathbf{T}_{\mathrm{final}} = \argmin_{\mathbf{T}}(1.0 - \mathcal{S}_e),
\end{equation}

\noindent where $\mathbf{T}_{\mathrm{final}}$ represents the final estimated transform. The minimization was achieved using the Covariance Matrix Adaptation Evolution Strategy (CMA-ES) optimization method \cite{hansen2003}.

The initial transform $\mathbf{T}_{\mathrm{init}}$ was manually estimated using the model-video overlay tool from the C3VD code repository. The CMA-ES optimization then refined this initial estimate. The optimization used a population size of 100. The parameter search space for the transformation was bounded within $\pm$0.1 radians for rotations and $\pm$7.5 millimeters for translations relative to the initial transform. Additionally, the initial step size $\sigma$ was set to 0.1. In some cases, the search space and sigma were further reduced for finer alignment.

The 2D/3D registration algorithm leveraged the similarity of edge features between video keyframes and rendered depth maps derived from the 3D model. Edges in the depth maps predominantly correspond to anatomical landmarks, such as haustral folds, which are often partially occluded by debris in the RGB frames. To mitigate this limitation, video sequences v2 and v3 were acquired from the same phantom before and after debris, using an identical robotic camera trajectory. Given the stationary phantom orientation between acquisitions and sub-millimeter repeatability of the robotic arm, the model transform calculated during v2 registration was reused for the debris-affected v3 sequence. This approach facilitated consistent model-video alignment despite partial visual obstruction.

To synchronize the clean (v2) and debris-filled (v3) colonoscopy videos, the temporal offset maximizing the correlation between their motion trajectories was calculated. Translation components were extracted from the camera pose logs of both videos, and their correlation was computed by sliding one trajectory over the other. The offset yielding the highest correlation was used to align the videos. Excess frames were then trimmed, resulting in paired frames of clean and debris-filled colon from the two videos.

\section*{Data Records}
The dataset is hosted on the Johns Hopkins Research Data Repository at \href{https://doi.org/10.7281/T1/JC64MK}{https://doi.org/10.7281/T1/JC64MK}, and is also available for preview and download at the project webpage - \href{https://durrlab.github.io/C3VDv2/}{https://durrlab.github.io/C3VDv2/}. The dataset comprises two distinct colon shapes (c1 and c2), each segmented into seven to eight anatomical regions, with each segment further characterized by four unique textures and colors (t1, t2, t3, and t4). A total of eight full colon phantom models were created from sixty phantom segments.

C3VDv2 contains 192 videos with a total of 169,371 frames. It comprises three different types of video sequences:
\begin{itemize}
    \item \textbf{Pixel-level GT Videos:} \textit{registered\_videos} were acquired with a static, undeformed colon phantom and are provided with per-frame ground truth maps (depth, normals, optical flow, etc.). Up to three videos were recorded per phantom segment:
    \begin{itemize}
        \item \textbf{v1:} clean colon with a baseline camera trajectory and imaging settings.
        \item \textbf{v2:} clean colon with a different camera trajectory and imaging settings as v1.
        \item \textbf{v3:} debris-filled colon using the same camera trajectory and imaging settings as v2.
    \end{itemize}
    This category includes 169 short videos with a total of 67,886 frames.
    \item \textbf{Deformation Videos:} \textit{deformation\_videos} consist of \textbf{v4} videos featuring externally induced active phantom deformation, captured with either static or linear camera motion. All videos include debris. Each folder contains all recorded RGB frames and a corresponding \textit{pose.txt} file (if camera is not stationary). The camera poses are in a frame-wise homogeneous format. This folder contains 15 short videos with a total of 6,185 frames.
    \item \textbf{Simulated Screening Videos:} \textit{simulated\_screening\_videos} comprise full-colon withdrawal sequences performed by a gastroenterologist to capture realistic camera motion. Similar to deformation videos, only RGB frames and camera poses in \textit{pose.txt} are provided. A total of 8 videos are included, comprising 95,300 frames.
\end{itemize}

Pixel-level ground truth was generated only for the rigid phantom videos (v1–v3) in the \textit{registered\_videos} folder. The folder structure for each registered video is as follows:

\begin{itemize}
	\item \textbf{RGB Frame:} \textit{rgb/NNNN.png} represents the raw (distorted, uncompressed) video frame from the Olympus CF-HQ190L video colonoscope. The black border with video metadata was cropped, resulting in an image size of 1350 x 1080 pixels. \textit{NNNN} denotes the 4-digit frame number within the video.
	\item \textbf{Depth Frame:} \textit{depth/NNNN\_depth.tiff} represents the depth along the camera frame's Z-axis, clamped between 0 and 100 mm, and linearly scaled and stored as a 16-bit grayscale image. For example, a pixel value of 16,384 corresponds to a depth of 25 mm. 
	\item \textbf{Surface Normal Frame:} \textit{normals/NNNN\_normals.tiff} stores the X, Y, and Z components of the surface normal vector for each surfel in the R, G, and B color channels, respectively. Components are linearly scaled from $\pm$1 to 0-65535 and stored as a 16-bit color image. Normal vector directions are defined with respect to the camera coordinate system: +x points right, +y points down, and +z points into the camera.
	\item \textbf{Optical Flow Frame:} \textit{optical\_flow/NNNN\_flow.tiff} depicts the optical flow from the current to the previous frame. X-direction motion (left to right, clamped between -20 to 20 pixels) is stored in the red channel, and Y-direction motion (up to down, clamped between -20 to 20 pixels) is stored in the green channel. Flow values are linearly scaled and stored as a 16-bit color image.
	\item \textbf{Occlusion Frame:} \textit{occlusions/NNNN\_occlusion.tiff} indicates pixels that occlude other mesh faces within 100 mm of the camera origin, assigning a value of 255 to these pixels and 0 to all others. This binary data is stored as an 8-bit grayscale image.
	\item \textbf{Diffuse Frame:} \textit{diffuse/NNNN\_diffuse.png} encodes Lambertian reflectance, computed using the dot product of the surface normal and the direction of the incident light. Reflectance values range from 0.1 to 1.0 and are linearly scaled and stored as an 8-bit grayscale image.
	\item \textbf{Camera Pose:} \textit{pose.txt} contains each frame's flattened homogeneous camera-to-world transformation matrix (row major order).
	\item \textbf{3D Model and Coverage Map:} \textit{coverage\_mesh.obj} stores the ground truth triangulated mesh. Texture vertices (vt) store coverage values, where vt=1 indicates an observed face, and vt=2 indicates an unobserved face.
\end{itemize}

During video recording, parameters such as camera trajectory, speed, and edge enhancement settings were varied to simulate diverse colonoscopy scenarios. These variations, along with simulated artifacts and challenging cases, are comprehensively documented in the \textit{C3VDv2\_Data\_Summary\_Sheet.xlsx}. It contains the following columns:

\begin{itemize}
	\item \textbf{Nomenclature:} The video name follows the format \textit{colon\_segment\_texture\_videonumber}, corresponding to the colon shape, anatomical segment, phantom texture, and video number.
	\item \textbf{Relative Camera Speed:} This refers to the percentage scaling factor (0-100\%) of the UR3e arm's maximum programmed velocity during motion segments. Actual trajectory velocities are non-uniform due to automatic acceleration/deceleration between waypoints.
	\item \textbf{Edge Enhancement:} This post-processing image enhancement mode increases image sharpness according to the selected setting, ranging from 1 (smoothest) to 3 (sharpest).
	\item \textbf{Debris:} This is a binary value of 0 for images of a clean colon and 1 for images of a colon with debris.
	\item \textbf{Deformation:} This is a binary value with 1 indicating deformation was externally applied to the phantom during image acquisition and 0 for no deformation. This is true for v4 videos.
	\item \textbf{Open End Visible:} Colon segment phantoms have two open ends, one of which is used for camera insertion. This binary value indicates 1 if the other end is visible in the video and 0 if the other end is not visible.
	\item \textbf{Tags:} Contains a list of scenarios occurring in the video. The tags indicate the presence of `polyps', `saturation', `water on lens', `debris on lens', `water jet', and camera motion and trajectories such as  `fast', `straight line', `loop', `helical', `mirrored path' (second half of the camera trajectory is the reverse of the first half), `zigzag', and `textureless surface enface'.
	\item \textbf{Comments:} Includes additional information such as the type of debris (e.g. blood and fecal), camera motion (e.g. enface to down-the-barrel transition), events in videos (e.g. polyp cleaning with water jet). 
	\item \textbf{Qualitative Score:} Subjective score from 1 to 3 indicating the quality of alignment between RGB frames and rendered views from the 3D model. 1 - Best Alignment, 2 - Good Alignment, 3 - Misalignment due to phantom manufacturing defects. 
	\item \textbf{Quantitative Score:} Comprises a combination of dice score (overlap) and chamfer distance (misalignment) on edges extracted from RGB frames and depth frames rendered from the 3D model. Refer to the technical validation section for more details on this score.
    \item \textbf{YouTube Preview URL ID:} Provides the ID to append to https://www.youtube.com/watch?v= to access a preview of the video containing the RGB frames, depth, normal, optical flow maps, and camera trajectory.
    \item \textbf{Total Frames:} Indicates the total number of frames in the video folder.
\end{itemize}

The spherical omnidirectional camera intrinsics are given in \textit{camera\_intrinsics.txt}. Additionally, two calibration sequences are provided for geometric and photometric calibration in the \textit{camera\_calibration} folder:
\begin{itemize}
	\item \textbf{camera\_calib\_checker.avi:} This video captures a $8\times11$ checkerboard pattern with a square size of 10 mm.
	\item \textbf{camera\_calib\_vicalib.avi:} This video features the ``big\_pattern'' target from the Endomapper repository \cite{azagra2023endomapper}. The \textit{big\_pattern.pdf} file, when printed at 100\% scale, includes a grid spacing of 5.29 mm, a large radius of 1.58 mm, and a small radius of 1.06 mm.
\end{itemize} 

\section*{Technical Validation}

\subsubsection*{Clean and Debris-Filled Colon Video Synchronization} 
C3VDv2 contains paired clean colon (v2) and debris-filled colon videos (v3), with identical camera trajectories and acquisition settings. To synchronize the videos, the time delay was calculated by maximizing the correlation between the translation components of the robotic arm poses. After trimming the beginnings and/or ends of the videos to account for this delay, the model transform from the clean colon video was applied to the debris-filled colon video. This transform reuse relies on two assumptions: the relative orientation of the model to the robotic arm remained unchanged between the two acquisitions, and the robotic arm has sub-millimeter pose repeatability. To validate these assumptions and assess frame alignment, the similarity between paired frames was assessed using the Structural Similarity Index (SSIM). Ten randomly selected frames from the clean colon video were each compared against a 31-frame window (offsets from +15 to -15) centered on the corresponding debris-filled colon frame. The average SSIM over the 10 frames was reported for each offset. The plot of average SSIM and standard deviation across all C3DVv2 video pairs (figure \ref{fig:technical_validation}a) shows a maximum at an offset of 0 frames (SSIM = 0.91), indicating maximum similarity between paired images. Notably, the synchronization method utilizes information only from the camera poses, whereas the validation method relies solely on comparing frames from the videos. This cross-modal validation using independent data sources confirms the accuracy of the synchronization method and justifies reusing the model transform for debris-filled videos.

\subsubsection*{2D/3D Registration Scoring}
An edge-based 2D/3D registration method was utilized to align the 3D model with video frames by optimizing the model transformation. The registration algorithm was previously validated in C3VD \cite{bobrow2023colonoscopy} using synthetic sequences. The proposed loss function demonstrated lower rotational and translational error compared to alternative loss functions.

While the registration performed well across most videos, some showed suboptimal alignment between the model and the real phantom, primarily due to phantom manufacturing defects. Since the true transformation for real videos is unknown, a qualitative scoring system was implemented to assess the perceived alignment between the video and the 3D model. The qualitative alignment assessment was performed by manually reviewing video frames overlaid with their corresponding depth maps rendered from the 3D model.

Videos were rated on a scale of 1 to 3. Score 1 indicates the best alignment. Score 2 represents good alignment, where local phantom defects lead to minor misalignment. Score 3 indicates a more considerable misalignment due to manufacturing defects in the phantom.

Most Score 3 cases stemmed from silicone bubbles during casting. We note that pouring silicone into a mold sometimes results in bubbles partially embedded in the casted silicone surface. After curing, extra silicone was applied to patch these bubble holes; however, this can lead to an overflow outside the holes. Consequently, when such a phantom was placed in the mold for imaging, any excess silicone would cause local deformation of the phantom shape. 

Additionally, quantitative scoring was performed by evaluating the alignment of edges extracted from the 3D model and the RGB frames from the videos. For each clean colon video, edges were extracted from both rendered depth frames and raw RGB frames using the DexiNed model. These edges were binarized, and the Dice coefficient and the Chamfer distance between the edge maps were computed and averaged across all video frames. Quantitative alignment scores $Q_{\text{align}}$ were computed by combining standardized z-scores of the Dice coefficient $D_z$ (edge similarity) and Chamfer distance $C_z$ (edge displacement). The final metric is defined as:

\begin{equation}
Q_{\text{align}} = \frac{1}{2} \left( D_z - C_z \right). \label{eq:composite_metric}
\end{equation}

A higher value of the quantitative alignment metric signifies better alignment. This formulation rewards high Dice scores (strong overlap) and penalizes high Chamfer distances (large misalignment). The Dice coefficient is insensitive to the degree of misalignment when edges do not overlap; hence, both the Dice coefficient and the Chamfer distance are used.

\begin{figure}[ht]
\centering
\includegraphics[width=\linewidth]{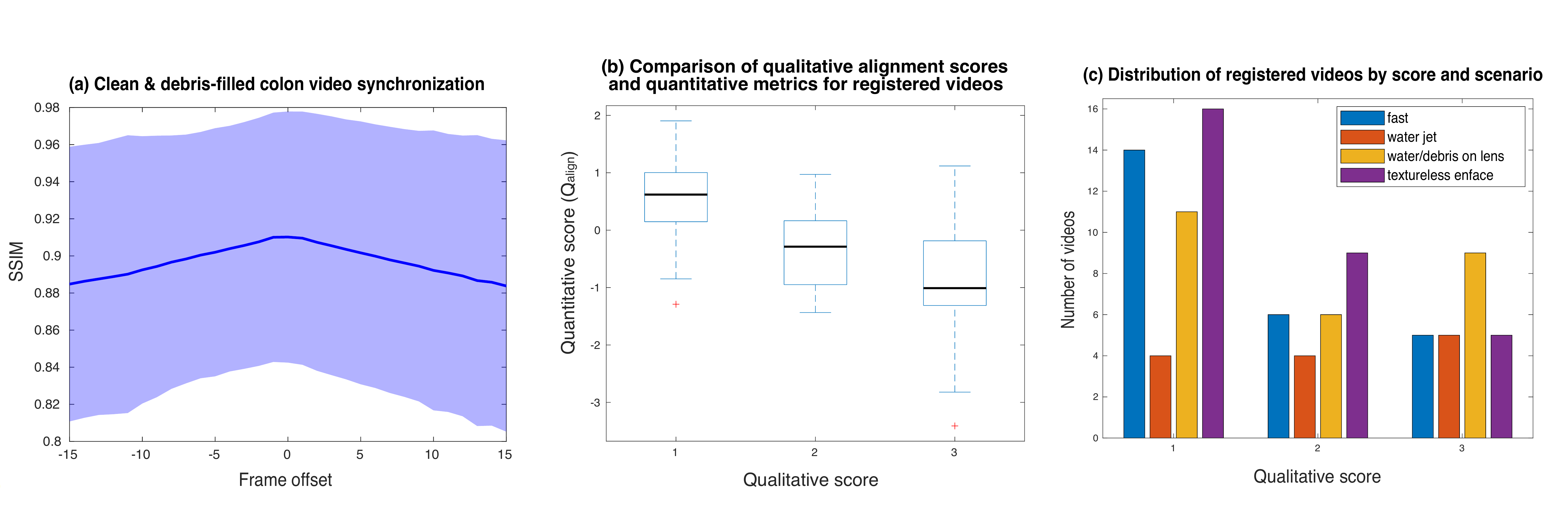}
\caption{\textbf{Technical validation results:} \textbf{(a)} To verify the synchronization between the clean (v2) and debris-filled (v3) videos, the structural similarity index (SSIM) was computed across a window of frames with offsets. Maximum SSIM at offset 0 indicates that clean \& debris-filled colon videos are well synchronized. \textbf{(b)} The alignment of the RGB video frames and the depth maps was qualitatively and quantitatively evaluated. The quantitative score comprises a combination of dice score and chamfer distance, while the qualitative score was given by visual inspection of RGB video frames overlaid on rendered depth maps (1 - best, 3 - least). The qualitative scores agree with the quantitative scores. \textbf{(c)} Histogram of challenging scenarios with respect to qualitative score. This indicates the presence of realistic scenarios, even in well-aligned videos with scores of 1 \& 2.}
\label{fig:technical_validation}
\end{figure}

A comparative analysis of each video's quantitative and qualitative metrics revealed a strong correlation, as illustrated in figure \ref{fig:technical_validation}b. We observed that most videos classified with a qualitative score of 1 exhibited higher quantitative alignment scores compared to those classified as 3. However, some Score 1 videos demonstrated low quantitative scores, often due to the presence of very few edges. In such cases, a minor misalignment of the few existing edges can disproportionately lower the score, even when the overall registration is qualitatively excellent. Conversely, videos classified as Score 2 or 3, which exhibited high quantitative scores, generally maintained good alignment for most edges. However, they were categorized as Score 2 and 3 due to phantom deformations, which affected only a small subset of edges, thus allowing the quantitative score to remain high. Figure \ref{fig:technical_validation}c shows the distribution of challenging scenarios across different scores.

\section*{Usage Notes}
It is important to note that all pixel-wise ground truth maps and 3D models for debris-filled colon videos (v3) are derived from the corresponding `clean' version of the colon without debris. These do not account for changes in depth, normals, or occlusion due to the addition of debris. The deformation videos (v4) and simulated screening videos do not include pixel-wise ground truth. For tasks requiring high pixel-level accuracy, such as depth estimation or normal prediction, videos with scores 1 and 2 are recommended. Score 3 videos may be useful for training weakly supervised, semi-supervised, or self-supervised paradigms \cite{tian2024endoomni} in addition to pose estimation. 

All pixel-wise ground truth was acquired from rigid, static colon models, which deviates from the deformable in~vivo colon environment. Real colonic tissue has dense vasculature \cite{golhar2018blood, golhar2017robust}, and polyps have microscopic surface patterns, which are not replicated in the phantoms used in the dataset. These features may serve as key points for feature-tracking algorithms in real colonoscopy, but may fail in the case of phantoms.

\section*{Data Availability}
The dataset is available for download and preview on the project webpage \href{https://durrlab.github.io/C3VDv2/}{https://durrlab.github.io/C3VDv2/}. It is hosted publicly on the Johns Hopkins Research Data Repository at \href{https://doi.org/10.7281/T1/JC64MK}{https://doi.org/10.7281/T1/JC64MK} and released under the CC BY-NC-SA 4.0 license. 

\section*{Code Availability}
The code used for the dataset is publicly available at \href{https://github.com/DurrLab/C3VD}{https://github.com/DurrLab/C3VD}.


\section*{Acknowledgments} 
This work was supported in part with funding and products provided by Olympus Corporation of the Americas. Although the agreement states a Sponsored Research Agreement, Olympus is funding, but not sponsoring this research. The authors would like to thank Lauren Shepard, Jinchi Wei, Dr. Ahmed Ghazi, and Dr. Ali Uneri of the Carnegie Center for Surgical Innovation, Dr. Swaroop Vedula of Malone Center for Engineering in Healthcare, and Dr. Jianing Li, and Dr. Surya Evani of the Division of Gastroenterology \& Hepatology at the Johns Hopkins Hospital for their assistance and for providing resources for data collection. The authors would also like to thank researchers at the University of North Carolina at Chapel Hill, University College London, and the University of Zaragoza for their feedback on the original C3VD dataset.

\section*{Author contributions statement}
M.V.G., T.L.B., and N.J.D. conceptualized and designed the study. M.V.G., L.S.G.F., and L.A. fabricated the phantoms, acquired and processed the data. T.L.B. implemented the registration and rendering pipeline. V.S.A. provided clinical oversight and performed simulated colonoscopies. M.V.G. performed the analysis and technical validation. M.V.G., L.A., L.S.G.F., and N.J.D. wrote the initial manuscript draft. All authors contributed to manuscript review and editing. N.J.D. supervised the project and acquired funding.

\section*{Competing interests}
V.S.A. is the co-founder of Origin Endoscopy Inc., Sotelix Endoscopy Inc., Flow Therapeutics Inc. and Solv Endotherapy Inc., and a consultant for Dragonfly Endoscopy, Ionis Pharmacy, and Olympus Medical. V.S.A has educational grants from Boston Scientific and Medtronic, and research grants from Nestle and Pentax Medical. All other authors have no relevant competing interests.

\section*{Figures \& Tables}

\begin{table*}[htbp]
\centering

\caption{\label{tab:comparisonMethods}Comparison of existing monocular colonoscopy datasets}
\footnotesize
\setlength\tabcolsep{1.5 pt}
\resizebox{\textwidth}{!}{
\begin{tabular}{l l l l l c c c c c c c r l}
\toprule
\textbf{\makecell{Camera \\ type}} & \textbf{Res.} & \textbf{FOV} & \textbf{Tissue Type} & \textbf{\makecell{3D GT\\Source}}  & \textbf{Pose} & \textbf{\makecell{Provided\\3D GT}} & \textbf{Debris} & \textbf{Blood} & \textbf{Foam} & \makecell{\textbf{Water} \\ \textbf{on Lens}} & \textbf{Deformation} &  \textbf{Frames} & \textbf{Dataset} \\
\midrule
USB & SD  &  Narrow    &  Phantom         & None     & \checkmark  & None & & & & & \checkmark & 23,935 & Fulton et al \cite{fulton2020}          \\
Rendered & SD  &  Narrow    & Virtual         & CT       &             & Depth  & & & & &  & 16,016 & Rau et al \cite{rau2019}             \\
Rendered &  SD      &   Narrow  &  Virtual         & CT                & \checkmark  & Depth & & & & & & 37,800& SimCol3D \cite{rau2023bimodal, rau2024simcol3d}   \\
Rendered &    SD/HD   & Narrow &  Virtual          &  CT                &  & None & &  \checkmark & & & & 14,181 & Dongdong et al \cite{he2025synthesized}    \\
USB &   SD/HD  & Wide   &  Ex-vivo Porcine & OS                 & \checkmark  &  3D Model  & & & & & & 39,406 &EndoSLAM\cite{ozyoruk2020}         \\
Pill Cam &  SD    & Wide   & Ex-vivo Porcine & None                & \checkmark  &  None & & & & & & 3,294 & EndoSLAM \cite{ozyoruk2020}         \\
Colonoscope &   HD   & Wide   & Phantom         & CT              &  \checkmark  & 3D Model &  & & & & & 12,250  & EndoSLAM \cite{ozyoruk2020}         \\
Rendered &   SD   & Narrow   & Virtual         & CT              &  \checkmark  & Depth & & & & & & 21,887  & EndoSLAM \cite{ozyoruk2020}         \\
Colonoscope &   HD    & Wide & In vivo Human        & None              &  \checkmark  & None & \checkmark & \checkmark & \checkmark  & \checkmark  & \checkmark  & 24 hrs video  &EndoMapper \cite{azagra2023endomapper}         \\
Colonoscope &   HD &  Wide & Phantom & Sculpted &  \checkmark & \makecell{Depth, Normals, \\Flow, 3D Model }& & & & & & 10,015 & C3VD\cite{bobrow2023colonoscopy} \\
\midrule
Colonoscope &   HD &  Wide & Phantom & Sculpted  &  \checkmark & \makecell{Depth, Normals, \\Flow, 3D Model, Diffuse} & \checkmark & \checkmark & \checkmark & \checkmark & & 67,886 & \makecell[l]{\textbf{C3VDv2} \\ (Pixel-wise GT videos)}    \\
Colonoscope &   HD &  Wide & Phantom &  Sculpted  &  \checkmark & \makecell{Undeformed\\3D Model} & \checkmark & \checkmark & \checkmark & \checkmark & \checkmark & 6,185 & \makecell[l]{\textbf{C3VDv2} \\ (Deformation videos)}  \\
Colonoscope &   HD &  Wide & Phantom &  Sculpted  &  \checkmark & 3D model & \checkmark & \checkmark & \checkmark & \checkmark &  & 95,300 & \makecell[l]{\textbf{C3VDv2} \\ (Sim. Screening videos)}  \\
\bottomrule
\end{tabular}
}
\scriptsize{Ground Truth (GT), Standard Definition (SD), High Definition (HD), Computed Tomography (CT), Optical Scan (OS).}
\end{table*}



\begin{table}[]
\caption{Ground truth 3D colon model attributes for Colon 1}
\centering
\label{tab:colon1-stats}
\resizebox{\textwidth}{!}{
\begin{tabular}{lllllll}
\hline
\textbf{Segment} &\textbf{Location}      & \textbf{Length (mm)} & \textbf{Object number} & \textbf{Type}                                                             & \textbf{Paris Classification}                                      & \textbf{Major Axis (mm)}                                                                     \\ \hline
\multirow{3}{*}{A} & \multirow{3}{*}{Cecum} & \multirow{3}{*}{125}& 1 & Adenoma & Ip & 2.8 \& 1.4\\ 
 &  &  & 2 & Hyperplastic & Is & 4.8\\ 
 &  &  & 3 & Ileocecal Valve & - & 24.8\\ \hline
B & Ascending              & 164& -     & -                                                                            & -                                                                  & -                                                                                            \\ \hline
C & Transverse 2           & 132& 4       & Adenoma                                                                          & IIa + Is                                                          & 10.2                                                                                           \\ \hline
D & Transverse 1           & 160& -     & -                                                                            & -                                                                  & -                                                                                            \\ \hline
E & Descending             & 149&  & & & \\ \hline
F & Sigmoid 1              & 150& 5& Adenoma& Is& 6.7\\ \hline
G & Sigmoid 2              & 108& -        & -                                                                        & -                                                                  & -                                                                                            \\ \hline
H & Rectum                 & 167& -           & -                                                                     & -                                                                  & -                                                                                            \\ \hline
\end{tabular}
}
\end{table}

\begin{table}[]
\caption{Ground truth 3D colon model attributes for Colon 2}
\centering
\label{tab:colon2-stats}
\resizebox{\textwidth}{!}{
\begin{tabular}{@{}lllllll@{}}
\toprule
\textbf{Segment}   & \textbf{Location}          & \textbf{Length (mm)} & \textbf{Object number}                                              & \textbf{Type}                                                                              & \textbf{Paris Classification}                                           & \textbf{Major Axis (mm)}                                                    \\ \midrule
A                  & Cecum                      & 83                   & 1                                                                   & Ileocecal valve                                                                            & -                                                                       & 15                                                                          \\ \midrule
\multirow{3}{*}{B} & \multirow{3}{*}{Ascending} & \multirow{3}{*}{160} & 2 & Adenoma & Ip & 5.7 \\
&  &  & 3 & Adenoma & Is & 11.4\\ 
 &  &  & 4 & Hyperplastic & IIa & 6.2\\ \hline
C                  & Transverse 1               & 148                  & 5                                                                   & Hyperplastic                                                                               & IIa                                                                     & 6.0                                                                        \\ \midrule
D                  & Transverse 2               & 100                  & 6                                                                   & Adenoma                                                                                    & Is                                                                      & 4.7                                                                         \\ \midrule
E                  & Descending                 & 170                  & -                                                                   & -                                                                                          & -                                                                       & -                                                                           \\ \midrule
F & Sigmoid   & 94  & 7                                                  & Adenoma                                                                  & Is                                                     & 3.6                                                        \\ \hline
G                  & Rectum                     & 105                  & 8                                                                   & Hyperplastic                                                                               & Is                                                                     & 11.6                                                                        \\ \bottomrule
\end{tabular}
}
\end{table}

\end{document}